\documentclass[runningheads]{llncs}
\usepackage[T1]{fontenc}
\usepackage{graphicx}
\usepackage{booktabs}
\usepackage{caption}
\usepackage{subcaption}
\usepackage{amsmath,amssymb,amsfonts}
\usepackage{multirow}
\usepackage{hyperref}

\begin{document}
\title{Dissecting RISC-V Performance: Practical PMU Profiling and Hardware-Agnostic Roofline Analysis on Emerging Platforms}
%
%
%
\authorrunning{A. Batashev}

\titlerunning{Practical PMU and Roofline on Emerging Platforms}

%

\author{Alexander Batashev\inst{1}\orcidID{0009-0003-0349-1415}}

\institute{Department of High-Performance Computing and Systems Programming, N. I. Lobachevsky State University of Nizhny Novgorod, Russia}

\maketitle              
\begin{abstract}
As RISC-V architectures proliferate across embedded and high-performance domains, developers face
persistent challenges in performance optimization due to fragmented tooling, immature hardware features, and
platform-specific defects. This paper delivers a pragmatic methodology for extracting actionable performance
insights on RISC-V systems, even under constrained or unreliable hardware conditions. We present a workaround
to circumvent hardware bugs in one of the popular RISC-V implementations, enabling robust event sampling. For
memory-compute bottleneck analysis, we introduce compiler-driven Roofline tooling that operates without
hardware PMU dependencies, leveraging LLVM-based instrumentation to derive operational intensity and
throughput metrics directly from application IR. Our open source toolchain automates these workarounds,
unifying PMU data correction and compiler-guided Roofline construction into a single workflow.

\end{abstract}

\section{Introduction}\label{sec1}

The RISC-V instruction set architecture (ISA) is rapidly gaining traction
across the computing landscape, from deeply embedded systems to
high-performance computing clusters. Its open nature, modularity, and
potential for customization offer significant advantages. However, as with
any emerging hardware ecosystem, achieving optimal performance on diverse
RISC-V implementations presents considerable hurdles for software developers
and system architects.

While performance analysis is crucial for optimization, the current RISC-V
ecosystem often suffers from fragmented tooling, immature hardware features,
and platform-specific defects, particularly concerning Performance Monitoring
Units (PMUs) . Unlike established architectures with mature profiling tools
and relatively consistent PMU behavior, developers targeting RISC-V frequently
encounter unreliable hardware counters, incomplete kernel support, or outright
hardware bugs that prevent standard profiling techniques like event sampling.
This lack of robust, reliable performance observability complicates effective
bottleneck analysis and optimization efforts, potentially limiting the
adoption and performance potential of RISC-V platforms.

This paper addresses these challenges by presenting a pragmatic methodology
for extracting actionable performance insights on RISC-V systems, even under
constrained or unreliable hardware conditions. We focus on practical
workarounds and hardware-agnostic techniques that lower the barrier to
effective performance analysis. Our approach combines techniques to robustly
utilize available PMU features with a novel compiler-driven method for
performance modeling that bypasses direct reliance on hardware counters.

The key contributions of this work are threefold: 
\begin{enumerate}
\item \textbf{A Practical PMU Sampling Workaround:} We identify and demonstrate
a technique to enable reliable event sampling for crucial metrics (like cycles
and instructions needed for IPC) on specific RISC-V hardware (SpacemiT X60) where
standard mechanisms fail due to hardware limitations, leveraging observed interactions
within the Linux \textit{perf\_event} subsystem.
\item \textbf{Hardware-Agnostic Roofline Analysis:} We introduce a compiler-driven Roofline
modeling approach using LLVM-based instrumentation. This method derives operational intensity
and throughput metrics directly from the application's Intermediate Representation (IR),
eliminating the dependency on hardware PMU counters often required by traditional Roofline tools
and ensuring applicability across diverse or limited RISC-V hardware.
\item \textbf{An Integrated Open-Source Toolchain:} We provide an open-source toolset that automates
these techniques, unifying the PMU sampling workaround (\textit{miniperf}) and the compiler-guided Roofline
construction into a practical workflow for RISC-V performance analysis.
\end{enumerate}

\section{Related work}\label{related}

Optimizing application performance on modern processors presents significant challenges due to increasing
microarchitectural complexity, workload diversity, and variations across hardware implementations.
This complexity makes identifying performance bottlenecks a difficult task. Researchers and industry have
developed various methodologies and tools to help developers in this process.

\subsection{Performance Monitoring Unit counters}\label{related_pmu}

Performance Monitoring Units (PMUs) are a cornerstone in understanding hardware execution characteristics.
Modern high-end processors often expose hundreds of performance events, but interpreting this vast amount of data
to pinpoint actual bottlenecks remains challenging. To address this, Intel Labs researchers pioneered Top-Down
Analysis method, aiming to hierarchically identify bottlenecks in complex out-of-order processors using a minimal
set of specific performance events\cite{yasin2014top}. This method simplifies analysis, reducing the steep learning curve associated
with microarchitectural details, and has been adopted in production by tools like Intel VTune.

SiFive researchers attempted to build an approximation of the TMA method for their hardware\cite{mou2024top}. Although some of their
results are applicable to many existing RISC-V implementations, the work is mostly specific to SiFive.

In our work, we will show some of the techniques we used to close the gaps in one of the more accessible implementations
of the RISC-V architecture. Although not in line with the full capabilities of TMA, it provides a solid foundation for
future research.

\subsection{Performance modeling and Roofline Analysis}

Beyond direct PMU counter analysis, performance models provide intuitive insights into application limitations. The Roofline model,
in particular, is noted for its straightforward guidance on whether an application is memory-bound or compute-bound on a specific system\cite{lo2015roofline}.
Constructing these models typically involves benchmarking system capabilities (peak performance, memory bandwidths) and measuring application
performance characteristics (throughput, operational intensity), often relying heavily on hardware PMU counters. The need for tools that
automate model construction and application analysis across different architectures is well-recognized.

The cache-aware Roofline model (CARM) tool provides such automation. CARM includes micro-benchmarks for assessing key performance characteristics
of target hardware, as well as support for dynamic binary instrumentation for extracting application arithmetic intensity and memory usage\cite{morgado2024carm}.

Although we find the methodology solid for usage on established platforms, initial enabling on emerging ones requires
significant investments for tools development. Our work diverges by proposing a hardware-agnostic Roofline approach.
By leveraging compiler instrumentation (specifically LLVM IR), we derive operational intensity and performance metrics
without direct dependence on PMU counters, offering a viable analysis path even on hardware with limited or faulty
monitoring capabilities.

\section{Accessing PMU counters on RISC-V hardware}\label{pmu_access}

Leveraging hardware Performance Monitoring Units (PMUs) is essential for in-depth performance analysis.
On Linux systems, the standard interface for accessing PMUs is the perf tool and its
underlying kernel subsystem, \textit{perf\_event}.

The RISC-V ISA provides a standardized framework for performance monitoring through the Sscofpmf extension
(Supervisor and Counter Overflow/Filtering for Performance Monitoring Facility), enabling cycle/instruction
counters and hardware event sampling. However, practical access to these features requires coordinated efforts
across the Linux kernel, OpenSBI firmware, and user-space tools like perf\cite{domingos2021supporting}.

\subsection{Architectural support}

The RISC-V Privileged Specification defines a standard set of Control and Status Registers (CSRs) for performance
monitoring. The core registers include: 

\begin{itemize}
\item \textit{mcycle:} Machine cycle counter (counts processor clock cycles)
\item \textit{minstret:} Machine instructions-retired counter (counts completed instructions).
\item \textit{mhpmcounter[3-31]:} Machine Hardware Performance Monitor Counters. These are generic counters
available for tracking various microarchitectural events. The number of available mhpmcounter registers is
implementation-defined.
\item \textit{mhpmevent[3-31]:} Machine Hardware Performance Monitor Event Selectors. Each corresponds to an
mhpmcounter register and is programmed with a specific event code (defined by the hardware vendor) to select
what the counter should track.
\item \textit{mcountinhibit:} A control register used to enable or disable the mcycle, minstret, and
mhpmcounter registers globally or individually.
\end{itemize}

Despite providing plenty of general-purpose registers, the specific events that these counters
can be configured to measure via the \textit{mhpmevent} registers are not standardized; they are
explicitly defined as platform- or implementation-specific. While future specification versions might
introduce standardization for common ISA-level or micro-architectural events, such as cache misses or specific
instruction types, the current ratified versions leave event definition beyond cycles and instructions retired
up to the hardware implementer.

\subsection{Software support in Linux}

The perf user-space tool relies on the kernel's \textit{perf\_event} subsystem.
When perf initiates monitoring (e.g., perf stat, perf record), it uses the
\textit{perf\_event\_open()} system call. This syscall requests the kernel to configure
specific performance events (hardware or software) for counting or sampling. The kernel's
architecture-specific PMU drivers are then responsible for: 

\begin{itemize}
\item \textbf{Programming the Hardware:} Configuring the PMU control registers to count the requested events.
\item \textbf{Managing Counters:} Enabling, disabling, reading, and resetting the hardware counters.
\item \textbf{Handling Overflows:} If sampling is requested, configuring the PMU to generate an interrupt when
a counter overflows. The interrupt handler collects context information (PC, registers, call stack) and
records the sample.
\item \textbf{Data Transfer:} Making counter values or collected samples available to the user-space perf
tool, typically via efficient ring buffers mapped into the application's address space.
\end{itemize}

\begin{figure}[h]
\centering
\includegraphics[width=0.63\textwidth]{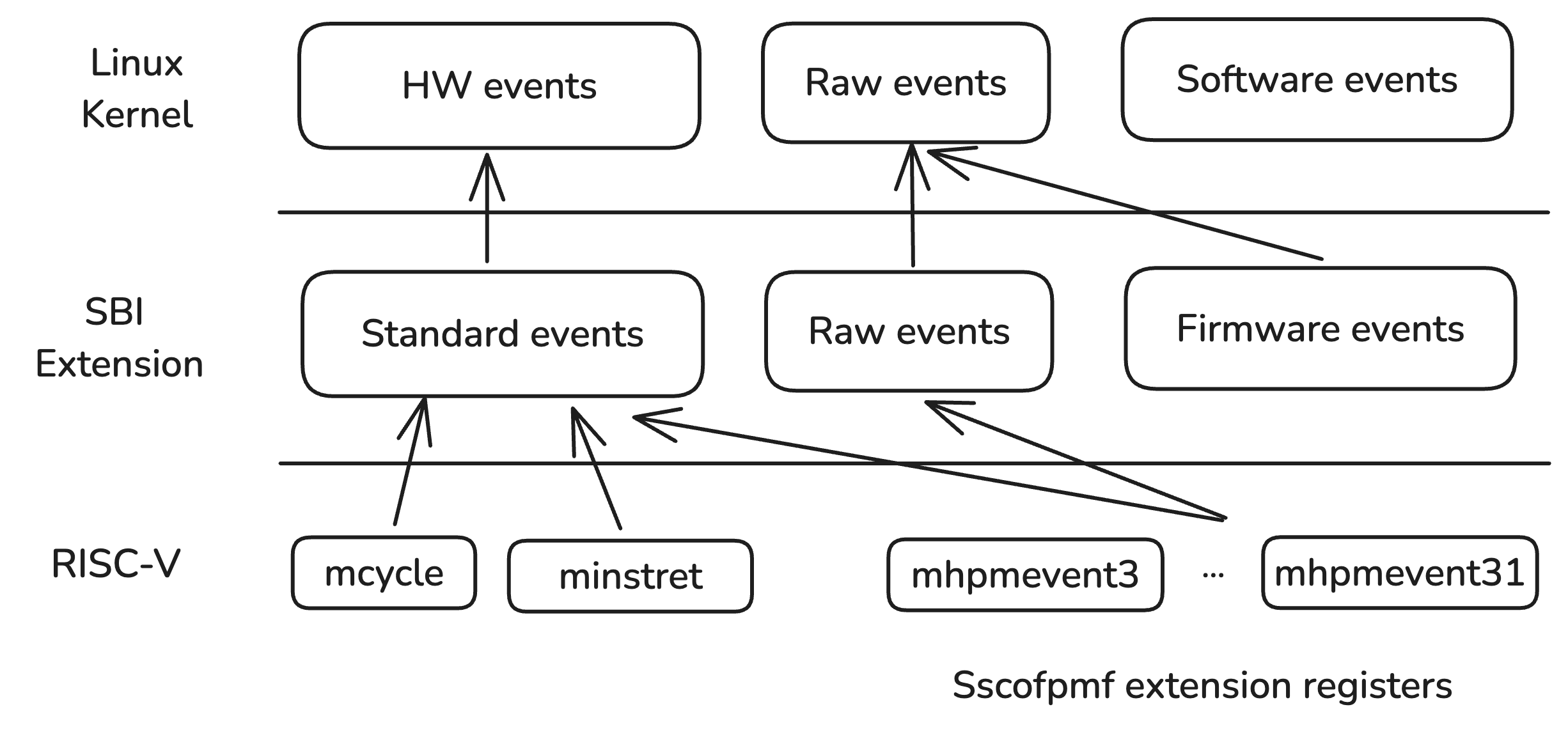}
\caption{Architecture of PMU counters software layer}\label{fig:linux_perf}
\end{figure}

The Linux kernel, operating in Supervisor mode, typically lacks the necessary privileges to directly configure
or access machine-level Performance Monitoring Unit (PMU) registers, such as the event configuration registers
(\textit{mhpmevent\#}) or the counter inhibit register (\textit{mcountinhibit}). To bridge this privilege gap,
communication with a Machine-mode entity like OpenSBI is required. This approach utilizes a dedicated
OpenSBI Hardware Performance Monitoring (HPM) extension, which defines specific functions callable via the
standard RISC-V environmental call (ecall) mechanism. Through this SBI extension, the kernel driver can
request OpenSBI to perform privileged read and write operations on its behalf, targeting machine-level PMU
registers like. Furthermore, to optimize performance monitoring and reduce overhead, the kernel can leverage
an SBI call to configure the mcounteren register, thereby enabling direct read access to HPM counters from
Supervisor mode, avoiding repeated SBI calls for counter reads. Figure \ref{fig:linux_perf} demonstrates the
relation between the RISC-V architecture and system software layers.

\subsection{Hardware support and current limitations}
While the previous sections outlined the theoretical framework for accessing PMU
counters through Linux's perf subsystem and the SBI interface, the practical
implementation of these mechanisms varies significantly across RISC-V hardware.
The specification-compliant interfaces described above represent an ideal
scenario, but the reality of current RISC-V implementations often diverges from
this ideal. As the RISC-V ecosystem continues to evolve, hardware vendors have
implemented PMU capabilities with varying degrees of completeness and compliance
with the specifications . These inconsistencies directly impact the
effectiveness of performance analysis tools and methodologies.

To establish a comprehensive understanding of performance analysis challenges in
the RISC-V ecosystem, we conducted a systematic evaluation of commercially 
available RISC-V platforms accessible to researchers and the broader developer 
community. Our investigation focused on three representative cores that span 
different microarchitectural approaches and feature sets: the SiFive U74,
T-Head C910, and SpacemiT X60. These cores were selected based on their 
availability in consumer-grade development boards (VisionFive II, Lichee Pi 4A, 
and Banana Pi F3 and Milk-V Jupyter, respectively), making our findings relevant 
to a wide range of potential users. 

Our comparative analysis, summarized in Table \ref{tab:riscv_hw_caps}, examines 
several critical dimensions that impact performance analysis capabilities: 

\begin{enumerate}     
\item \textbf{Microarchitectural organization:} We assessed whether each core
implements an in-order or out-of-order execution model, which significantly
affects both performance characteristics and the complexity of performance
analysis. 
\item \textbf{Vector extension support:} The presence and version of the RISC-V
Vector (RVV) extension were evaluated, as this capability is crucial for
high-performance computing workloads and introduces additional performance
monitoring considerations. 
\item \textbf{PMU counter capabilities:} We specifically examined support for
counter overflow interrupts, which are essential for sampling-based profiling
techniques. This feature varies significantly across implementations, from 
non-existent to fully supported. 
\item \textbf{Upstream kernel support:} The degree of integration into
mainstream Linux repositories was assessed, as this affects the accessibility
and stability of performance monitoring tools. 
\end{enumerate}

\begin{table}[b]
\centering
\caption{Comparison of available RISC-V hardware capabilities}
\label{tab:riscv_hw_caps}
\begin{tabular}{|l|c|c|c|}
\hline
Core                       & \multicolumn{1}{c|}{SiFive U74} & \multicolumn{1}{c|}{T-Head C910} & \multicolumn{1}{c|}{SpacemiT X60} \\ \hline
Out-of-Order               & No                              & Yes                              & No                                \\ \hline
RVV version                & Not suppoorted                  & 0.7.1                            & 1.0                               \\ \hline
Overflow interrupt support & No                              & Yes                              & Limited                           \\ \hline
Upstream Linux support     & Yes                             & Partial                          & No                                \\ \hline
\end{tabular}
\end{table}

Our analysis reveals significant heterogeneity across these platforms. The
T-Head C910 offers the most advanced microarchitecture with out-of-order
execution and comprehensive PMU support, but relies heavily on vendor-specific
kernel modifications. The SpacemiT X60, while featuring the latest RVV 1.0
specification, provides only limited PMU sampling capabilities through
non-standard counters. The SiFive U74, despite better upstream Linux
integration, lacks both vector extensions and overflow interrupt support,
severely limiting traditional performance analysis approaches. 

\subsubsection{PMU sampling on X60}

Although the SpacemiT X60 core lacks hardware support for overflow interrupts
on the standard \textit{mcycle} and \textit{minstret} counters, an examination
of vendor-provided kernel source code revealed the presence of three non-standard counters that do support sampling: \textit{u\_mode\_cycle},
\textit{m\_mode\_cycle}, and \textit{s\_mode\_cycle}. These counters track
cycles spent in User, Machine, and Supervisor modes, respectively, though 
corresponding instruction counters are unfortunately absent.

During experimentation with \textit{perf\_event\_open()} system call on a Spacemit X60 system,
we observed that configuring one of these sampling-capable counters as
a leader group causes \textit{mcycles} and \textit{minstret} to be sampled
concurrently within that group, triggered by the leader's overflow frequency.

To leverage this observed behavior, we developed \textit{miniperf}\footnote{\url{https://github.com/alexbatashev/miniperf}}, a tool
wrapping the standard \textit{perf\_event\_open()} API.
Unlike the standard \textit{perf} utility, it
automatically groups counters and selects an appropriate sampling-capable
leader. This technique enables the collection of crucial performance metrics,
such as Instructions Per Cycle (IPC), despite the documented lack of direct
sampling support for the necessary underlying counters
\footnote{\url{https://bianbu.spacemit.com/en/development/perf/}}. Furthermore, \textit{miniperf} adopts a distinct approach
to hardware compatibility; rather than utilizing standard perf event discovery mechanisms, it relies solely on
CPU identification registers. This direct hardware identification enables more robust management of supported
features and platform-specific workarounds. 

\section{Compiler-based approach for Roofline analysis}\label{roofline}

\subsection{LLVM Background}

Before diving into our specific implementation, it's helpful to understand some
key LLVM concepts that form the foundation of our approach. LLVM is a compiler
infrastructure that provides a collection of modular compiler and toolchain
technologies. Two concepts are particularly relevant to our work: 

\textbf{LLVM Intermediate Representation (IR)} is a platform-independent code 
representation that sits between the source code and machine code in the
compilation process\cite{lattner2004llvm}. LLVM IR serves as a common format throughout the
compilation pipeline, allowing optimizations to be applied regardless of source 
language or target architecture. This IR uses a RISC-like instruction set but 
includes higher-level information such as types, explicit control flow graphs, 
and memory operations. For performance analysis, LLVM IR provides an ideal 
observation point as it retains enough high-level structure to identify loops and functions,
exposes memory operations and arithmetic instructions explicitly
and emains independent of the target architecture's specific instruction set.

\textbf{LLVM Passes} are transformations or analyses that operate on the code as
it moves through the compilation pipeline\cite{lattner2004llvm}. Passes can analyze the code or modify it. The pass infrastructure allows for
modular compiler extensions without modifying the core compiler. Common passes
include loop analysis, dead code elimination, and instruction combining. Our
approach implements a custom instrumentation pass that leverages existing 
analysis passes to identify and instrument performance-critical regions. 

\subsection{Compiler instrumentation details}
Our approach implements a Clang compiler plugin that adds instrumentation to
collect metrics needed for Roofline analysis without relying on PMU counters.

The instrumentation process consists of several key steps: 

\begin{enumerate}
\item \textbf{Loop Nest Identification:} Our pass first traverses each
function in the program, using LLVM's Loop Analysis infrastructure to
identify loop nests. Loop nests are particularly important for performance
analysis as they often contain the computational kernels that dominate
execution time.

\item \textbf{Region Extraction:} For each identified loop nest, we use
LLVM's RegionInfoAnalysis to ensure the region has a single entry and single
exit point (SESE). This property is crucial for clean extraction and
instrumentation. The CodeExtractor utility then outlines this region into a
separate function.

\item \textbf{Function Duplication:} The extracted function is cloned to create two versions: 
the original (unmodified) function and an instrumented version that collects performance metrics.

\item \textbf{Call Site Modification:} The original call site is modified to
include logic that selects between the two function versions based on 
environment variables. This allows for runtime control over which regions 
are instrumented. The call site is also wrapped with special notification 
functions that signal the beginning and end of monitored regions:

\begin{verbatim}
LoopInfo LI{line=42, filename="foo.c", func_name="bar"};
LoopHandle *LH = mperf_roofline_internal_notify_loop_begin(LI);
if (mperf_roofline_internal_is_instrumented_profiling())
  bar_loop0_instrumented(args..., LH);
else
  bar_loop0_outlined(args...);
mperf_roofline_internal_notify_loop_end(LH);
\end{verbatim}

\item \textbf{Metric Collection:} In the instrumented function version, we 
insert code at the basic block level to count bytes loaded to/from memory, 
integer arithmetic operations, and floating-point arithmetic operations.
\end{enumerate}

These counts are accumulated during execution and reported to our \textit{miniperf} tool via callback functions when the program terminates. 

\subsection{Runtime Analysis and Roofline Construction}

\begin{figure}[h]
\centering
\includegraphics[width=0.8\textwidth]{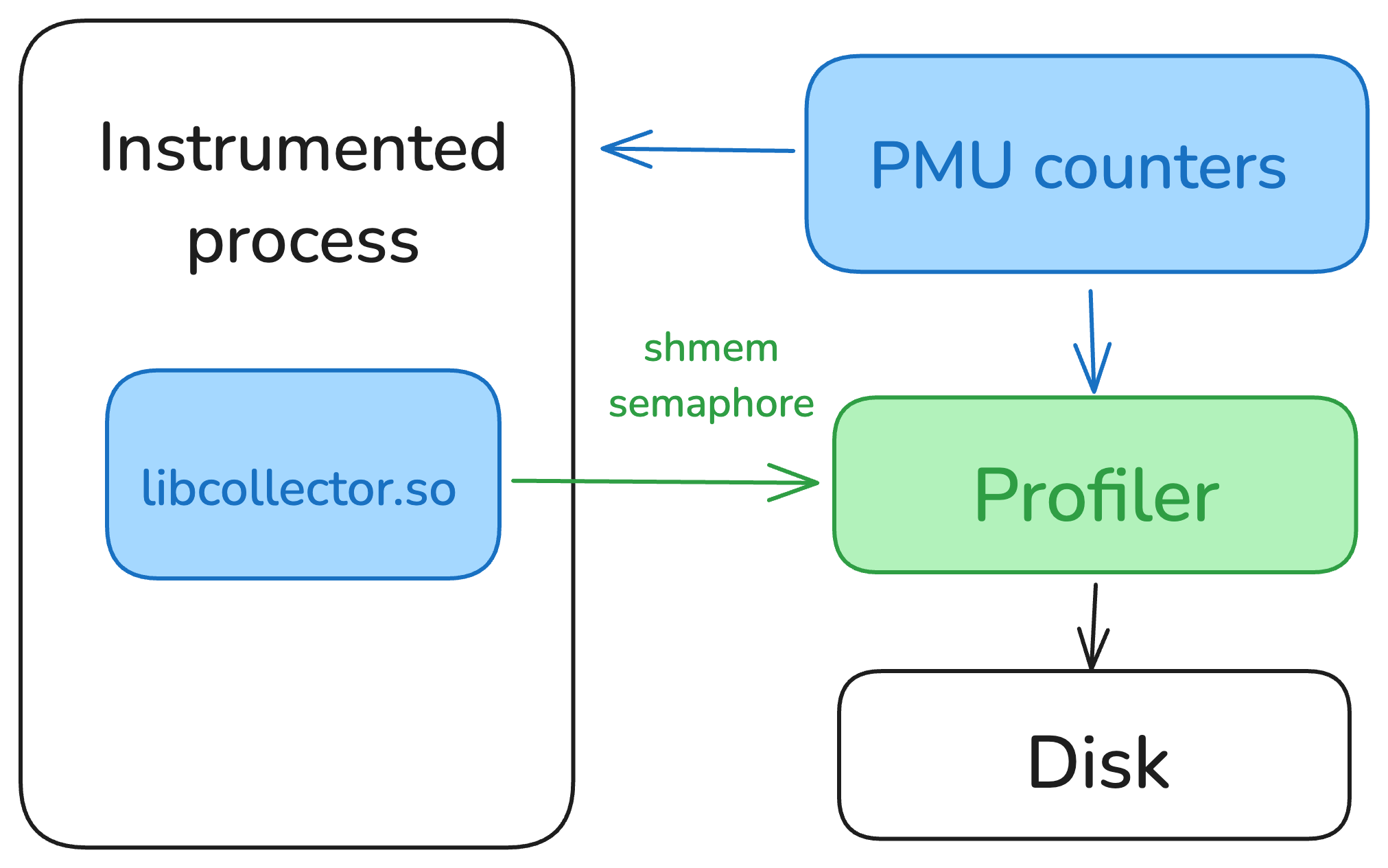}
\caption{Overview of instrumented workflow}\label{fig:comp_instr}
\end{figure}

The actual Roofline model construction happens through a two-phase execution
approach (illustrated by figure \ref{fig:comp_instr}): 
\begin{enumerate}
\item \textbf{Baseline Execution:} The program runs with instrumentation disabled to establish baseline performance. 
\item \textbf{Instrumented Execution:} The program runs again with instrumentation enabled for targeted regions.
\end{enumerate}

Our tool coordinates these executions and correlates the results, calculating: 
\begin{itemize}
\item Execution time differences between instrumented and non-instrumented runs
\item Memory traffic (bytes/second) derived from load/store counts
\item Computational throughput (operations/second)
\item Arithmetic intensity (operations/byte)
\end{itemize}

These metrics can be then plotted against the hardware capabilities (peak
compute performance and memory bandwidth) to construct the Roofline model.
Since our approach derives these metrics from LLVM IR rather than
hardware events, the analysis remains consistent across RISC-V
implementations.

\subsection{Advantages and limitations}
This compiler-based approach offers several advantages for performance analysis: 
\begin{itemize}
\item \textbf{Hardware Independence:} By operating at the IR level, our approach 
works across different hardware implementations regardless of PMU support or 
specific extensions. Adding a new platform is a matter of writing a good LLVM backend
for the new target, which is done as part of a regular compiler development cycle.
\item \textbf{Fine-grained Analysis:} We can target specific code regions rather
than whole-program profiling.
\item \textbf{Consistent Metrics:} The metrics are derived from program behavior
rather than hardware-specific events, providing a consistent view across 
platforms.
\end{itemize}

However, there are notable limitations: 
\begin{itemize}
\item \textbf{External Function Calls:} Loops containing calls to external 
functions (e.g., library calls) cannot be fully instrumented, as we cannot track 
operations inside those functions. This particularly affects code that relies 
on external libraries.
\item \textbf{Runtime Overhead:} The instrumentation adds significant overhead, 
making absolute performance measurements less accurate. This is mitigated by our 
two-phase execution approach.
\item \textbf{Compiler Optimizations:} Post-instrumentation optimizations might 
alter the relationship between IR-level metrics and actual hardware behavior. We 
address this by applying our pass late in the optimization pipeline.
\item \textbf{Deterministic Execution:} The approach assumes deterministic program behavior across runs, which may not hold for all applications.
\end{itemize}

Despite these limitations, our approach provides valuable insights into
application performance characteristics on RISC-V platforms where traditional
PMU-based analysis is unavailable or unreliable. The hardware-agnostic nature of
the approach makes it valuable in the rapidly evolving
RISC-V ecosystem. 

\section{Evaluation}\label{evaluation}

\subsection{Hotspot analysis}

Identifying CPU hotspots – the sections of code where the
processor spends most of its time – is a critical step in
performance optimization. While various profiling techniques
exist, Flame Graphs, invented by Brendan Gregg, offer an intuitive and powerful visualization for
understanding CPU usage patterns.

Flame Graphs are a visualization of profiled software,
created by sampling stack traces over a period. The x-axis of
a Flame Graph represents the stack profile population, with
stack frames sorted alphabetically to maximize merging,
rather than representing the passage of time. The y-axis
shows the stack depth, increasing from bottom to top. Each
rectangle in the graph corresponds to a function (a stack
frame). The width of a rectangle is directly proportional to
the frequency with which that function appeared in the
sampled stacks – wider frames indicate functions that were on-CPU
more often or were part of call stacks that were on-CPU more
often. The top edge of a frame indicates the function
currently executing on the CPU, while the frames below it
represent its parent callers (its ancestry). This
hierarchical visualization allows developers to quickly
pinpoint the most time-consuming code paths\cite{FlameGraph}.

The primary importance of Flame Graphs lies in their ability
to provide a quick, comprehensive, and deep view of CPU
usage. They distill potentially overwhelming amounts of
profiling data into a single, readable image, making it
easier to identify call paths that
consume the most CPU resources. This is invaluable for
guiding efforts toward the most impactful code areas.

\begin{figure}[h]
\centering
\begin{subfigure}[b]{0.9\textwidth}
    \includegraphics[width=\textwidth]{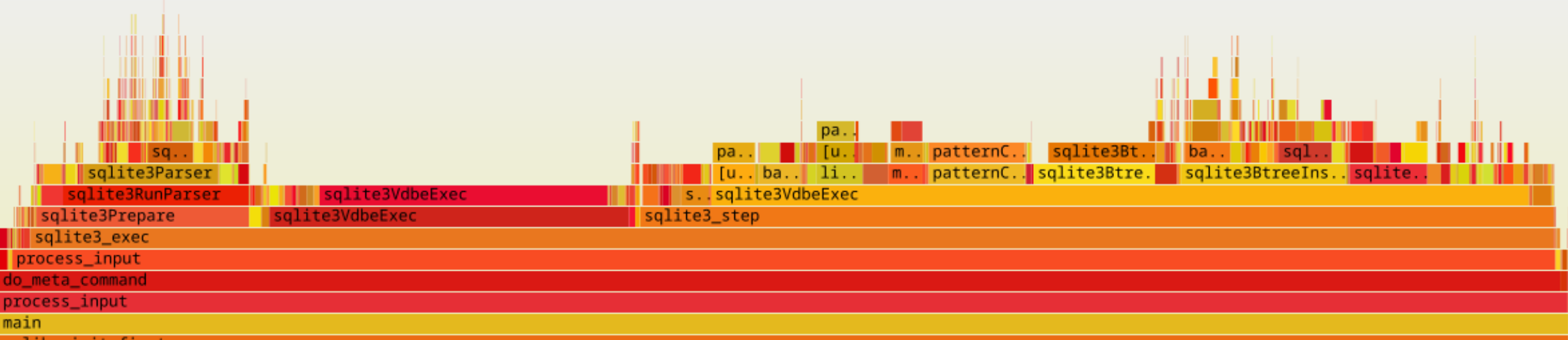}
    \caption{SpacemiT X60, cycles}
    \label{fig:x60_cycles}
\end{subfigure}
\begin{subfigure}[b]{0.9\textwidth}
    \includegraphics[width=\textwidth]{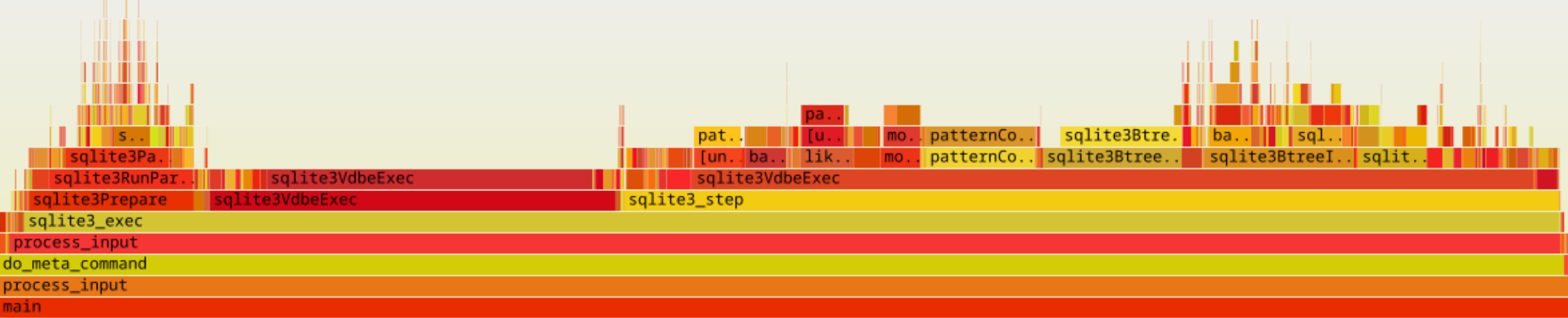}
    \caption{SpacemiT X60, instructions}
    \label{fig:x60_instructions}
\end{subfigure}
\begin{subfigure}[b]{0.9\textwidth}
    \includegraphics[width=\textwidth]{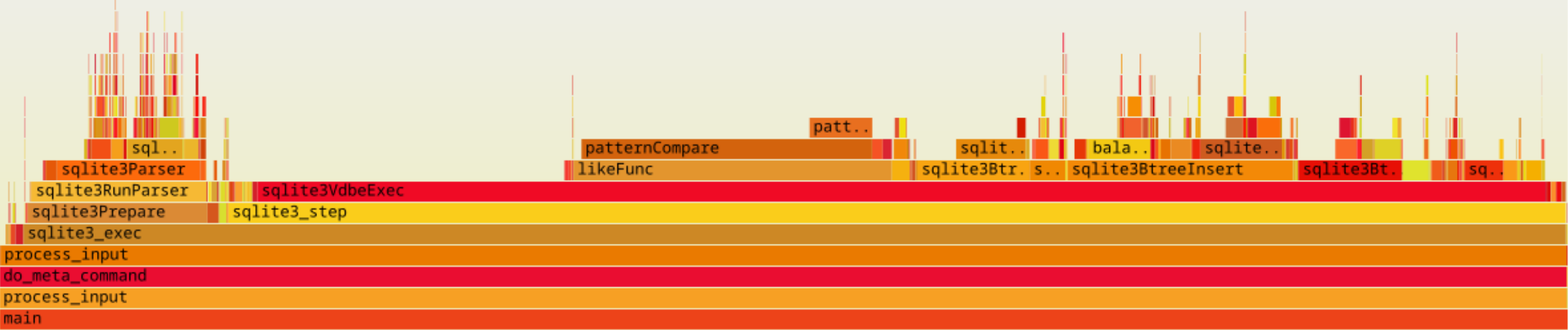}
    \caption{Intel Core i5-1135G7, cycles}
    \label{fig:x86_cycles}
\end{subfigure}
\begin{subfigure}[b]{0.9\textwidth}
    \includegraphics[width=\textwidth]{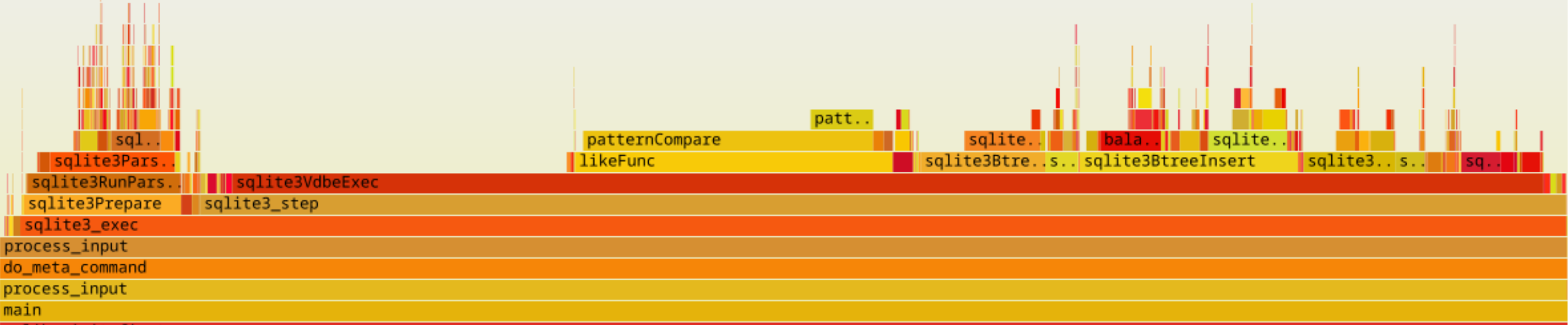}
    \caption{Intel Core i5-1135G7, instructions}
    \label{fig:x86_instructions}
\end{subfigure}
\caption{Flame Graphs for sqlite3 benchmark}
\label{fig:flamegraph}
\end{figure}

Our tool, \textit{miniperf}, facilitates the generation of Flame
Graphs using either CPU cycles or instructions retired as the
sampled metric. While cycle-based Flame Graphs directly
represent CPU time, Flame Graphs built from instructions
retired offer a valuable proxy metric, particularly useful
across diverse platforms for identifying code sections that
are underoptimized and warrant attention from programmers or
compiler developers. For instance, consider a function where
vectorization is expected; if the instructions retired Flame Graph shows
a significantly wider frame for this function
(e.g., processing 8x more scalar instructions than an
equivalent vectorized version would), it strongly suggests an
inferior vectorization scheme or a complete lack of
vectorization. Flame Graphs make such comparative analysis
visually intuitive - often as straightforward as comparing two
images. Crucially, \textit{miniperf}'s ability to enable sampling of cycles
and instructions retired, even on platforms with limited or
non-standard hardware support (as demonstrated with the
SpacemiT X60 workaround), empowers developers to perform
these vital optimizations and analyses ahead of mature
hardware platform releases, accelerating software readiness.

The results from our tool for the sqlite3 benchmark (taken from LLVM Test Suite),
depicted in the Flame Graphs (Figure \ref{fig:flamegraph}) and hotspot summary (Table \ref{tab:hotspots}),
clearly demonstrate the current performance landscape. While the x86 platform may
execute more instructions for the given task, its microarchitectural efficiency,
evidenced by an IPC reaching $3.38$, significantly surpasses that of the SpacemiT X60
core, which achieved an IPC of $0.86$. This highlights a considerable
performance gap and optimization opportunities on the RISC-V system.

\begin{table}[t]
\centering
\caption{Top 3 hotspots from sqlite3 benchmark}\label{tab:hotspots}
\begin{tabular}{|l|ccc|ccc|}
\hline
\multirow{2}{*}{Function} & \multicolumn{3}{c|}{SpacemiT X60}                                        & \multicolumn{3}{c|}{Intel Core i5-1135G7}                                 \\ \cline{2-7} 
                          & \multicolumn{1}{c|}{Total, \%} & \multicolumn{1}{c|}{Instructions} & IPC & \multicolumn{1}{c|}{Total, \%} & \multicolumn{1}{c|}{Instructions} & IPC \\ \hline
sqlite3VdbeExec           & \multicolumn{1}{r|}{$18.44\%$}         & \multicolumn{1}{l|}{$3,634,478,335$}            & $0.86$   & \multicolumn{1}{l|}{$19.58\%$}         & \multicolumn{1}{l|}{$6,737,784,530$}            & $3.38$   \\ \hline
patternCompare            & \multicolumn{1}{r|}{$11.63\%$}         & \multicolumn{1}{l|}{$2,298,438,217$}            & $0.86$   & \multicolumn{1}{l|}{$18.60\%$}         & \multicolumn{1}{l|}{$5,857,213,374$}            & $3.09$   \\ \hline
sqlite3BtreeParseCellPtr  & \multicolumn{1}{r|}{$10.17\%$}         & \multicolumn{1}{l|}{$1,905,893,304$}            & $0.82$   & \multicolumn{1}{l|}{$6.42\%$}         & \multicolumn{1}{l|}{$2,113,027,184$}            & $3.24$   \\ \hline
\end{tabular}
\end{table}

\subsection{Roofline analysis}

Beyond direct hotspot identification, understanding the interplay between a processor's
computational capabilities and its memory subsystem performance is crucial for
comprehensive optimization. The Roofline model provides an insightful visual
framework for this purpose, offering an easy way to characterize
application performance and identify primary bottlenecks. Key strengths of
the Roofline model include its ability to intuitively depict whether an
application is limited by memory bandwidth or by the processor's peak
computational throughput, thereby guiding optimization efforts effectively.
It correlates achieved performance (e.g., GFLOP/s) with arithmetic intensity
(operations per byte of memory accessed), providing clear upper bounds defined by
the hardware's capabilities. To demonstrate the practical application of our
compiler-driven Roofline analysis, we will utilize the following code kernel
for detailed examination:

\begin{verbatim}
for (int ii = 0; ii < n; ii += TILE_SIZE) {
  for (int jj = 0; jj < n; jj += TILE_SIZE)
    for (int kk = 0; kk < n; kk += TILE_SIZE)
      for (int i = ii; i < ii + TILE_SIZE && i < n; i++)
        for (int j = jj; j < jj + TILE_SIZE && j < n; j++) {
          float sum = C[i * n + j];
          for (int k = kk; k < kk + TILE_SIZE && k < n; k++)
            sum += A[i * n + k] * B[k * n + j];
          C[i * n + j] = sum;
        }
}
\end{verbatim}

For our experimental evaluation of the Roofline model, all benchmarks were compiled using Clang 19
with the \textit{-O3} optimization flag. For the x86 platform, we specifically enabled AVX2
instructions via the \textit{-mavx2} flag, while for RISC-V, we targeted the RV64GCV profile
using \textit{-march=rv64gcv}.

\begin{figure}[p]
\centering
\begin{subfigure}[b]{0.8\textwidth}
    \includegraphics[width=\textwidth]{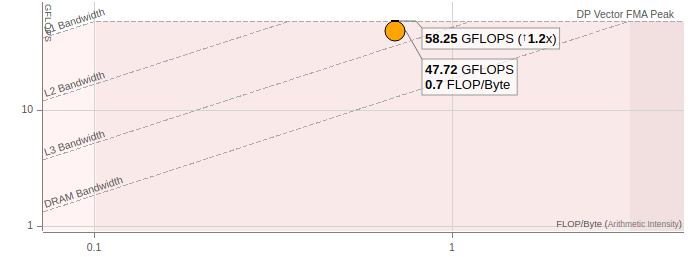}
    \caption{Intel Advisor Roofline model}
    \label{fig:x86_ia_roofline}
\end{subfigure}
\begin{subfigure}[b]{0.8\textwidth}
    \includegraphics[width=\textwidth]{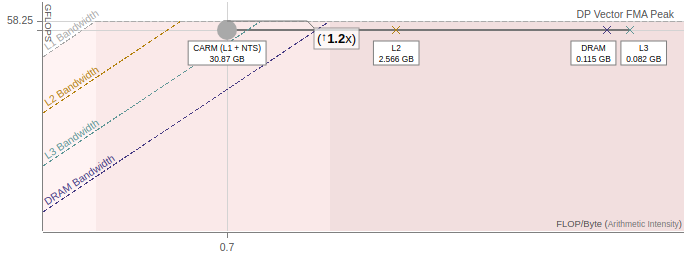}
    \caption{Intel Advisor Roofline model (CARM)}
    \label{fig:x86_carm_roofline}
\end{subfigure}
\begin{subfigure}[b]{0.8\textwidth}
    \includegraphics[width=\textwidth]{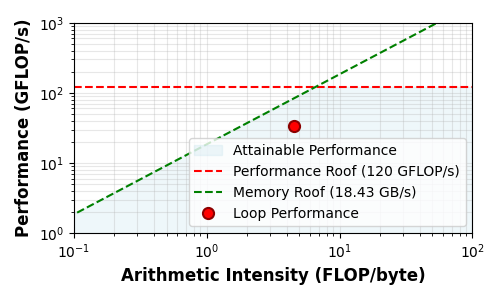}
    \caption{Intel Core i5-1135G7, miniperf Roofline}
    \label{fig:x86_miniperf_roofline}
\end{subfigure}
\begin{subfigure}[b]{0.8\textwidth}
    \includegraphics[width=\textwidth]{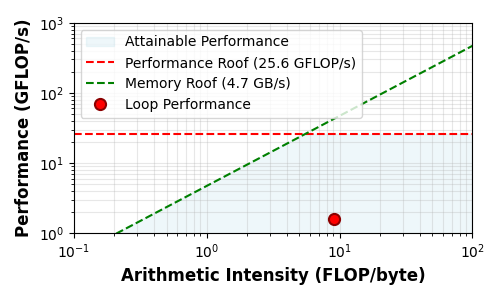}
    \caption{SpacemiT X60, miniperf Roofline}
    \label{fig:x60_miniperf_roofline}
\end{subfigure}
\caption{Roofline model for matmul kernel}
\label{fig:roofline}
\end{figure}

Figure \ref{fig:roofline} illustrates a comparative Roofline analysis for a representative kernel on an
x86 platform, contrasting the results obtained from our \textit{miniperf} tool with those from Intel Advisor. Our
compiler-driven instrumentation approach, as implemented in \textit{miniperf}, reported a performance of $34.06$
GFLOP/s for the kernel. This figure closely aligns with the benchmark's self-reported performance of $33.0$
GFLOP/s, indicating good fidelity with the application's own measurements. In contrast, Intel Advisor
reported a higher performance of $47.72$ GFLOP/s. This discrepancy primarily stems from the different
methodologies: \textit{miniperf} directly instruments the generated code at the LLVM IR level to count operations,
whereas Intel Advisor typically relies on dynamic analysis of hardware performance counters. The slight
variance between \textit{miniperf}'s measurement and the benchmark's self-reported value can be attributed to the
inherent overhead of our low-level loop instrumentation; any user-code measurements around the instrumented
loop will invariably include this instrumentation overhead, though, as observed, this impact is not
substantial. For the arithmetic intensity calculations presented, we currently focus on operations exposed to
the L1 cache. A more comprehensive memory hierarchy analysis would require deep, platform-specific knowledge,
which is challenging given the diverse and evolving landscape of RISC-V hardware. It is also worth
noting that for the x86 Roofline plot (Figure \ref{fig:roofline}), the performance ceilings (roofs) were
directly adopted from Intel Advisor, with our \textit{miniperf}-derived data point mapped onto this
established model.

Turning to the SpacemiT X60 RISC-V platform, we constructed its Roofline model using a combination of
established benchmarks and theoretical peaks due to the current limitations in obtaining precise, detailed
hardware specifications for all memory levels. For the memory bandwidth roof, we utilized results from Olaf
Bernstein's memset benchmark\footnote{\url{{https://camel-cdr.github.io/rvv-bench-results/bpi_f3/memset.html}}},
which indicates a peak performance of approximately $3.16$ bytes/cycle for this core. At a nominal frequency of
$1.6$ GHz, this translates to a theoretical DRAM memory bandwidth of roughly $4.7$ GB/s
($3.16\text{ bytes/cycle} * 1.6 \text{GHz}$). For the compute roof, we considered the theoretically
achievable peak: assuming 2 instructions retired per cycle (a common design target for in-order cores) and an
8-element single-precision floating-point vector capability (as per RVV 1.0 with a 256-bit VLEN), this yields a peak of $2\text{ IPC} * 8 \text{  SP FLOP/vector instruction} * 1.6 \text{GHz} = 25.6 \text{ GFLOP/s}$
(assuming one vector FMA or equivalent per cycle).

Against these theoretical ceilings, our kernel achieved $1.58$ GFLOP/s on the SpacemiT X60. This resulting
performance, significantly below the theoretical compute and memory bandwidth capabilities, indicates
substantial room for improvement. It highlights opportunities for compiler developers to enhance code
generation and vectorization for this specific core, for performance engineers to further tune applications,
and for hardware vendors to potentially refine microarchitectural efficiency or provide clearer performance
characterization. \textit{miniperf}'s ability to provide these hardware-agnostic operational metrics, even
when detailed hardware counter analysis is challenging, offers unique insights into such performance
characteristics and optimization potentials on emerging platforms.

This evaluation chapter successfully demonstrated the practical application of our proposed methodologies,
showcasing effective PMU profiling on limited RISC-V hardware and the utility of our compiler-driven,
hardware-agnostic Roofline analysis. Through hotspot analysis with Flame Graphs and comparative Roofline
modeling, our \textit{miniperf} tool effectively highlighted key performance characteristics and substantial
optimization opportunities. Ultimately, these evaluations confirm that our approach provides valuable,
actionable insights for developers navigating the complexities of the evolving RISC-V performance landscape,
even in the absence of mature hardware support.

\section{Conclusion}\label{conclusion}

The rapid proliferation of RISC-V architectures presents exciting opportunities
but also significant challenges for performance analysis and optimization. 
Developers often face a fragmented landscape characterized by inconsistent 
hardware features, particularly concerning Performance Monitoring Units (PMUs), 
and immature tooling support. This paper tackled these practical hurdles by 
introducing pragmatic methodologies designed to deliver actionable performance 
insights even on constrained or non-standard RISC-V platforms. 

Our key contributions provide tangible solutions for developers navigating this 
ecosystem. Firstly, we demonstrated a practical workaround for enabling crucial 
PMU event sampling (specifically \textit{mcycle} and \textit{minstret}) on the 
SpacemiT X60 platform, leveraging non-standard counters and Linux 
\textit{perf\_event} group behavior to overcome documented hardware limitations. 
This technique allows for the collection of fundamental metrics like IPC where 
it was previously thought impossible via standard sampling. Secondly, we 
introduced a novel, hardware-agnostic approach for Roofline performance 
modeling. By utilizing compiler-based instrumentation via an LLVM pass, this 
method derives operational intensity and throughput directly from the 
application's intermediate representation, circumventing the need for 
potentially unreliable or unavailable hardware PMU counters. Finally, we 
presented \textit{miniperf}, an open-source tool that integrates these techniques, 
automating the PMU sampling workaround and providing a unified interface for 
collecting data for our compiler-driven Roofline analysis, streamlining 
the performance analysis workflow on supported RISC-V systems.

While our current work provides valuable tools and techniques, several avenues 
exist for future enhancement and expansion. A primary focus will be on improving 
the usability and insightfulness of the \textit{miniperf} tool. Currently, performance 
data presentation is basic; we plan to incorporate more sophisticated 
visualization capabilities, including direct generation of Roofline 
plots within the tool and better correlation between PMU-derived metrics and 
compiler-level statistics. This will help users quickly interpret results 
and identify bottlenecks. 

Furthermore, we aim to significantly expand the analytical capabilities of \textit{miniperf}.
A key direction is the integration of structured performance analysis methodologies,
specifically the Top-Down Microarchitecture Analysis (TMA) method. Adapting TMA to RISC-V
requires careful mapping of its hierarchical bottleneck categories onto the available
PMU events across different RISC-V implementations. This presents a challenge given
the current PMU heterogeneity, but achieving even partial TMA support would provide
users with a much more systematic way to diagnose performance limitations beyond
the memory/compute focus of the Roofline model.

Beyond single-core microarchitectural analysis, we plan to extend \textit{miniperf}'s scope to
encompass parallel programming models. This involves incorporating support for
standardized tracing and tool interfaces such as OMPT (OpenMP Tools Interface)
and XPTI (oneAPI DPC++). The goal is to correlate low-level PMU data or
compiler-derived metrics collected by \textit{miniperf} with higher-level parallel
programming constructs, offering a more holistic view of application performance
and enabling the diagnosis of parallelism-specific bottlenecks.

Ultimately, this ongoing work aims to build a robust, accessible, and practical 
performance analysis toolkit that empowers developers to effectively optimize 
applications for the diverse and rapidly evolving RISC-V hardware landscape.


%
%
%
%
\bibliographystyle{splncs04}
\bibliography{bibliography}
\end{document}